\documentclass[showpacs]{revtex4}

\begin{document}

\title{Involute, minimal, outer and increasingly trapped spheres}

\author{Sean A. Hayward}
\affiliation{Center for Astrophysics, Shanghai Normal University, 100 Guilin 
Road, Shanghai 200234, China} 
\date{11th December 2009}

\begin{abstract}
Seven different refinements of trapped surfaces are proposed, each intended as 
potential stability conditions. This article concerns spherical symmetry, but 
each condition can be generalized. Involute trapped spheres satisfy a similar 
condition to minimal trapped spheres, which are are strictly minimal with 
respect to the Kodama vector. There is also a weaker version of involute 
trapped. Outer trapped spheres have positive surface gravity. Increasingly 
(future, respectively past) trapped spheres generate spheres which are more 
trapped in a (future, respectively past) causal direction, with three types: in 
any such causal direction, along the dual Kodama vector, and in some such 
causal direction. Assuming the null energy condition, the seven conditions form 
a strict hierarchy, in the above order. In static space-times, they reduce to 
three inequivalent definitions, namely minimal, outer and increasingly trapped 
spheres. For a widely considered class of so-called nice (or non-dirty) black 
holes, minimal trapped and outer trapped become equivalent. 
Reissner-Nordstr\"om black holes provide examples of this, and that 
increasingly trapped differs. Examples where all three refinements differ are 
provided by a simple family of dirty black holes parameterized by mass and 
singularity area. 
\end{abstract}
\maketitle 

\section{Introduction} 
Trapped surfaces, as originally defined by Penrose \cite{Pen}, play an 
important role in gravitational physics, both for black holes and in cosmology. 
Such surfaces were crucial in the singularity theorems of Penrose and Hawking 
\cite{Pen,Haw,HP}. More recent years have seen the development of a local, 
dynamical theory of black holes in terms of marginal surfaces, a limit of 
trapped surfaces, including laws of black-hole dynamics involving physical 
quantities such as mass and surface gravity \cite{AK, Boo, Kri, GJ,bh6}. 

However, a problem has become clear recently, as follows. Suppose one has a 
spherically symmetric space-time such as Vaidya, and considers the region 
covered by all trapped surfaces, not necessarily spherically symmetric. One 
might expect the boundary of this trapped region to consist of marginal 
surfaces, i.e.\ to be a trapping horizon. However, this is not so: trapped 
surfaces can poke through the trapping horizon \cite{SK,Ben,BS}. The boundary 
must be spherically symmetric, but can be outside the trapping horizon. On the 
other hand, this boundary does not have the special physical properties that 
trapping horizons have, such as a first law involving surface gravity 
\cite{1st} and a local Hawking temperature \cite{HCNVZ}. There is thus a 
conflict between the mathematics and physics, which is something of a crisis 
for anyone hoping to understand black holes in a practical way. 

In the author's view, the physics is clear and so the mathematics must yield. 
That is, trapped surfaces as simply defined need to be refined in some way to 
forbid the above behaviour. Of course, the condition or conditions should be 
geometrically natural, and preferably with some physical content. The situation 
appears to be similar to that which used to hold for marginal surfaces, which 
are too general to characterize black holes and needed to be refined, such as 
by the outer condition expressing positive surface gravity \cite{bh6,1st}, or 
other stability conditions \cite{New,KH,AMS,AMMS}. Such stability conditions 
are inequalities of one differential level higher than trapped or marginal 
itself. 

This article proposes seven such refinements, of essentially three types with 
some variations, each of which holds for any trapped sphere in a Schwarzschild 
space-time. They are of some interest in themselves and turn out to be related, 
assuming the Einstein equation, or more exactly just the null energy condition 
(NEC). A companion article will deal with general cases \cite{ts2}, while this 
article will be restricted to spherical symmetry, with everything respecting 
spherical symmetry. One reason for making this special study is that the 
physical meaning of all terms which will appear is clear, while in general one 
must use the available geometrical quantities, which are one step removed from 
familiar physical quantities, such as mass $m$, surface gravity $\kappa$ and 
energy flux $\psi$. Thus while the companion article \cite{ts2} is intended to 
be of more interest to mathematicians, this article is intended for those 
interested in more physical issues related to black holes which can be 
adequately addressed in spherical symmetry. 

The article is organized as follows. Section II explains how any sphere is 
extremal in some normal direction, and defines minimal trapped spheres. Section 
III defines outer trapped spheres and shows that the condition is implied by 
minimal trapped. Section IV defines increasingly trapped spheres of three kinds 
and shows that the conditions are implied by outer trapped. Section V defines 
involute trapped spheres of two kinds and shows that they are outer trapped.  
Section VI turns to static cases and shows that the conditions reduce to only 
three inequivalent ones, namely minimal, outer and increasingly trapped. 
Section VII studies so-called nice black holes, as opposed to so-called dirty 
black holes \cite{Vis}, and shows that minimal trapped and outer trapped become 
equivalent in such cases. Section VIII studies Reissner-Nordstr\"om black 
holes, finding the boundaries of the regions of the refined trapped spheres. 
Section IX does the same for a class of dirty black holes for which all three 
refinements differ. Section X concludes with hierarchy diagrams and remarks. 

\section{Minimal trapped spheres}
In terms of the area radius $r$, the area of the spheres of symmetry is
\begin{equation}
A=4\pi r^2.
\end{equation}
The dual Kodama vector is
\begin{equation}
k_*=g^{-1}(dr)
\end{equation}
where $g$ is the space-time metric. A sphere is said to be {\em untrapped, 
marginal} or {\em trapped} if $k_*$ is respectively spatial, null or temporal 
\cite{1st,sph,ine}. If the space-time is time-orientable and $k_*$ is future 
(respectively past) causal, then the sphere is said to be {\em future} 
(respectively {\em past}) trapped or marginal. 

There is a duality operation on normal vectors $\eta$, corresponding to the 
Hodge dual on 1-forms, defined by 
\begin{equation}
g(\eta_*,\eta)=0,\qquad g(\eta_*,\eta_*)=-g(\eta,\eta).
\end{equation}
Then $k$ is the Kodama vector \cite{1st,sph,Kod}, defined up to sign, which can 
be locally fixed so that $k$ is future-causal in untrapped regions. Then a 
sphere is trapped, marginal or untrapped if $k$ is respectively spatial, null 
or temporal. 

Now 
\begin{equation}
k\cdot dr=0
\end{equation}
which expresses that any sphere is extremal in the $k$ direction. In the 
special case $k=0$, the sphere is extremal in any normal direction, but 
otherwise $k$ gives the unique such direction. Thus a trapped sphere is 
equivalently defined as a sphere which is extremal in a unique spatial normal 
direction. Then it is natural to ask whether the sphere is not merely extremal 
but minimal, as is the case for any trapped sphere in a Schwarzschild 
space-time. 

Definition 1. A (strictly) {\em minimal} trapped sphere is a trapped sphere for 
which 
\begin{equation}\label{minimal}
k^ak^b\nabla_a\nabla_br>0
\end{equation}
where $\nabla$ is the covariant derivative operator of $g$. Note that 
minimality itself requires only a non-strict inequality, but the strict sign 
will turn out to be convenient. This definition has effectively been given by 
Maeda et al.\ in the context of cosmological wormholes \cite{MHC}. 

\section{Outer trapped spheres}
Surface gravity was defined as \cite{1st,ine,HCNVZ}  
\begin{equation}\label{kappa}
\kappa=\textstyle{\frac12}{*}d{*}dr
\end{equation}
where $d$ is the exterior derivative and $*$ the Hodge dual in the normal 
space, i.e.\ ${*}d{*}d$ is the normal wave operator, fixing the sign 
convention. Then a trapping horizon was said to be {\em outer, degenerate} or 
{\em inner} if respectively $\kappa>0$, $\kappa=0$ or $\kappa<0$. Examples of 
all types are provided by Reissner-Nordstr\"om solutions, where they correctly 
label the types of Killing horizon. This suggests extending the terminology to 
trapped spheres. 

Definition 2. An {\em outer} trapped sphere is a trapped sphere for which
\begin{equation}\label{outer}
\kappa>0.
\end{equation}
Therefore spheres sufficiently close to an outer trapping horizon will be outer 
trapped, while those sufficiently close to an inner trapping horizon will not, 
for instance in a non-degenerate Reissner-Nordstr\"om solution. A trapped 
sphere with $\kappa<0$ may similarly be called {\em inner} trapped. 

Lemma 1. Assuming the Einstein equation with units $G=1$, 
\begin{equation}\label{Q0}
k^ak^b\nabla_a\nabla_br=\left(\frac{2m}r-1\right)\kappa-4\pi rk_*\cdot\psi
\end{equation}
where $m$ is the mass \cite{1st,sph,ine,MS}, 
\begin{equation}\label{m0}
\frac{2m}r-1=g(k,k)
\end{equation}
and $\psi$ is an energy flux defined by \cite{1st,ine}
\begin{equation}\label{ps}
\psi=T\cdot k^*+wdr
\end{equation}
where $T$ is the energy tensor and $w$ is an energy density:
\begin{equation}\label{w}
w=-\textstyle{\frac12}\hbox{tr}\,T
\end{equation}
where the trace is in the normal space.

Proof. In dual-null coordinates $x^\pm$, the metric has the form
\begin{equation}
ds^2=r^2d\Omega^2-2e^{2\varphi}dx^+dx^-
\end{equation}
where $d\Omega^2$ refers to the unit sphere and $(r,\varphi)$ are functions of 
$(x^+,x^-)$. In these coordinates, trapped spheres are equivalently defined by 
$\partial_+r\partial_-r>0$, with $\partial_\pm r<0$ (respectively ${}>0$) for  
future (respectively past) trapped spheres, assuming that $\partial_\pm$ are 
future-pointing. Then one finds the explicit expressions
\begin{eqnarray}
&&dr=\partial_+rdx^++\partial_-rdx^-\\
&&k_*=-e^{-2\varphi}(\partial_+r\partial_-+\partial_-r\partial_+)\\
&&k=e^{-2\varphi}(\partial_+r\partial_--\partial_-r\partial_+)\\
&&2m/r-1=2e^{-2\varphi}\partial_+r\partial_-r\label{m}\\
&&\kappa=-e^{-2\varphi}\partial_+\partial_-r\label{kappa2}\\ 
&&w=e^{-2\varphi}T_{+-}\\
&&\psi=-e^{-2\varphi}(T_{++}\partial_-rdx^++T_{--}\partial_+rdx^-)\label{psi}
\end{eqnarray}
which implies 
\begin{equation}\label{k*psi}
k_*\cdot\psi=e^{-4\varphi}\left(T_{--}(\partial_+r)^2+T_{++}(\partial_-r)^2\right).
\end{equation}
The only relevant non-zero connection coefficients are 
$\Gamma^\pm_{\pm\pm}=2\partial_\pm\varphi$. Then
\begin{eqnarray}
k^ak^b\nabla_a\nabla_br&=&k^ak^b\left(\partial_a\partial_br-\Gamma^c_{ab}\partial_cr\right)\nonumber\\
&=&e^{-4\varphi}\left((\partial_-r)^2(\partial_+\partial_+r-2\partial_+\varphi\partial_+r) 
+(\partial_+r)^2(\partial_-\partial_-r-2\partial_-\varphi\partial_-r)
-2\partial_+r\partial_-r\partial_+\partial_-r\right).\label{Q1}
\end{eqnarray}
The null-null components of the Einstein equation are 
\begin{equation}\label{Ein}
\partial_\pm\partial_\pm 
r-2\partial_\pm\varphi\partial_\pm r=-4\pi rT_{\pm\pm}.
\end{equation}
Then straightforward calculation using (\ref{m}), (\ref{kappa2}), 
(\ref{k*psi}). 

Proposition 1. {\em NEC and minimal trapped implies outer trapped.}

Proof. NEC $\Rightarrow k_*\cdot\psi\ge0$, as is most easily seen from 
$T_{\pm\pm}\ge0$ and (\ref{k*psi}). For a trapped sphere, $r<2m$, then inspect 
signs in (\ref{minimal}), (\ref{outer}) and(\ref{Q0}).

\section{Increasingly trapped spheres}
Noting that $g(k,k)=2m/r-1$ (\ref{m0}) vanishes for marginal spheres and is 
positive for trapped spheres, it can be taken as a measure of how trapped a 
sphere is. The idea then is to ask whether it is increasing to the future 
(respectively past) for a future (respectively past) trapped sphere. 
Consideration of general cases \cite{ts2} suggests instead the measure 
$(2m/r-1)/r^2$, which yields stricter conditions. Three different definitions 
can be given, as follows. 

Definition 3. An {\em increasingly} trapped sphere is a trapped sphere for 
which 
\begin{equation}
k_*\cdot d\left(\frac1{r^2}\left(\frac{2m}r-1\right)\right)>0.
\end{equation}

Definition 4. An {\em anyhow increasingly} trapped sphere is a future 
(respectively past) trapped sphere for which, for all future (respectively 
past) causal normal vectors $\zeta$,
\begin{equation}
\zeta\cdot d\left(\frac1{r^2}\left(\frac{2m}r-1\right)\right)>0.
\end{equation}

Definition 5. A {\em somehow increasingly} trapped sphere is a future 
(respectively past) trapped sphere for which, for some future (respectively 
past) causal normal vector $\zeta$, 
\begin{equation}
\zeta\cdot d\left(\frac1{r^2}\left(\frac{2m}r-1\right)\right)>0.
\end{equation}

Clearly anyhow increasingly trapped implies increasingly trapped, which implies 
somehow increasingly trapped. 

Lemma 2. Assuming the Einstein equation, 
\begin{equation}\label{inc}
r^2\zeta\cdot d\left(\frac1{r^2}\left(\frac{2m}r-1\right)\right)=
8\pi r\zeta\cdot\psi-2\left(\kappa+\frac1r\left(\frac{2m}r-1\right)\right)\zeta\cdot dr.
\end{equation}

Proof. The last term above comes from the $1/r^2$ term, so it suffices to 
calculate
\begin{eqnarray}
\zeta\cdot d(m/r)&=&
(\zeta^+\partial_++\zeta^-\partial_-)(e^{-2\varphi}\partial_+r\partial_-r)\nonumber\\
&=&e^{-2\varphi}\left(\zeta^+\partial_-r(\partial_+\partial_+r-2\partial_+\varphi\partial_+r)
\zeta^-\partial_+r(\partial_-\partial_-r-2\partial_-\varphi\partial_-r) 
+(\zeta^+\partial_+r+\zeta^-\partial_-r)\partial_+\partial_-r\right)\nonumber\\
&=&4\pi r\zeta\cdot\psi-\kappa\zeta\cdot dr
\end{eqnarray}
using the Einstein equations (\ref{Ein}) as before and the expressions 
(\ref{kappa2}), (\ref{psi}). 

Proposition 2. {\em NEC and outer trapped implies anyhow increasingly trapped.}

Proof. $r<2m$, $\zeta\cdot dr<0$ for $\zeta$ in the given causal quadrant, and 
NEC $\Rightarrow \zeta\cdot\psi\ge0$, then inspect signs in (\ref{outer}), 
(\ref{inc}). 

If one wished to show directly that the sphere was merely increasingly trapped, 
the corresponding expression would be 
\begin{equation}
k_*\cdot d\left(\frac mr\right)=4\pi rk_*\cdot\psi+\left(\frac{2m}r-1\right)\kappa.
\end{equation}
This result will survive in general cases, while the above result will require 
a stricter definition of outer trapped \cite{ts2}, which in turn suggests a 
stricter condition than minimal trapped, described below. 

\section{Involute trapped spheres}
The minimality condition (\ref{minimal}) can be rewritten as
\begin{equation}
k_*^ak^b\nabla_bk_a<0
\end{equation}
due to orthogonality of $k_*$ and $k$. This form is useful for general 
calculations, and suggests further definitions. 

Definition 6. An {\em involute} trapped sphere is a future (respectively past) 
trapped sphere for which, for all future (respectively past) causal normal 
vectors $\zeta$, 
\begin{equation}\label{involute}
\zeta^ak^b\nabla_bk_a<0.
\end{equation}

Definition 7. A {\em somehow involute} trapped sphere is a future (respectively 
past) trapped sphere for which, for some future (respectively past) causal 
normal vector $\zeta$, 
\begin{equation}\label{sinvolute}
\zeta^ak^b\nabla_bk_a<0.
\end{equation}

Clearly involute trapped implies minimal trapped, which implies somehow 
involute trapped. 

Lemma 3. Assuming the Einstein equation, 
\begin{equation}\label{inv}
\zeta^ak^b\nabla_bk_a=\kappa\zeta\cdot dr+4\pi r\zeta\cdot\psi.
\end{equation}

Proof. 
\begin{eqnarray}
\zeta^ak^b\nabla_bk_a&=&\zeta^ak^b\left(\partial_bk_a-\Gamma^c_{ba}k_c\right)\nonumber\\
&=&e^{-2\varphi}\big(\zeta^+\partial_-r(\partial_+\partial_+r-2\partial_+\varphi\partial_+r) 
+\zeta^-\partial_+r(\partial_-\partial_-r-2\partial_-\varphi\partial_-r) 
-(\zeta^+\partial_+r+\zeta^-\partial_-r)\partial_+\partial_-r\big)\nonumber\\
&=&4\pi r\zeta\cdot\psi+\kappa\zeta\cdot dr
\end{eqnarray}
using the expressions (\ref{kappa2}), (\ref{psi}) and the Einstein equations 
(\ref{Ein}) as before. 

Proposition 3. {\em NEC and somehow involute trapped implies outer trapped.} 

Proof. As before, $\zeta\cdot dr<0$ and NEC $\Rightarrow \zeta\cdot\psi\ge0$, 
then inspect signs in (\ref{sinvolute}), (\ref{inv}). 

\section{Static space-times}
In static space-times, the static Killing vector $\partial_t$ is proportional 
to $k$. Thus $k\cdot df=0$ for any function $f$. 

Proposition 4. {\em In static space-times, the three increasingly trapped 
conditions 3--5 are equivalent.} 

Proof. Recall that $k$ and $k_*$ are orthogonal, and span the normal space for 
a trapped sphere. Any normal vector $\zeta$ can then be written as a linear 
combination of $k$ and $k_*$, with positive component along $k_*$ for the 
appropriate causal quadrant. Since
\begin{equation}
k\cdot d\left(\frac1{r^2}\left(\frac{2m}r-1\right)\right)=0
\end{equation}
the result follows. 

Proposition 5. {\em In static space-times, the minimal trapped condition 1 and 
the two involute trapped conditions 6--7 are equivalent.}

Proof. As above, since 
\begin{equation}
2k^ak^b\nabla_bk_a=k^b\nabla_b(k^ak_a)=0.
\end{equation}

Thus the seven (or eight in general) conditions reduce to three, Definitions 
1--3, which can be seen to be inequivalent in examples to follow. 

\section{Nice black holes}
Definition 8. A {\em nice} black hole is a static black hole for which 
\begin{equation}
g(\partial_t,\partial_t)=-g^{-1}(dr,dr)
\end{equation}
for some thereby standard choice of the scaling of the static Killing vector 
$\partial_t$. This includes Schwarzschild, Reissner-Nordstr\"om (de Sitter) and 
many other black holes which have been considered in the literature. A non-nice 
static black hole is here called a {\em dirty} black hole, more or less 
consistently with previous usage \cite{Vis}. 

For nice black holes, the Kodama and Killing vectors coincide, $k=\partial_t$, 
and the metric may be written, except at trapping horizons, as
\begin{equation}\label{nice}
ds^2=r^2d\Omega^2+\left(1-\frac{2m}r\right)^{-1}dr^2-\left(1-\frac{2m}r\right)dt^2.
\end{equation}
Note that this is valid both outside the horizon, $r>2m$, and inside, $r<2m$.

Proposition 6. {\em For nice black holes, minimal trapped and outer trapped are 
equivalent.}

Proof. For a static black hole, one finds the energy density (\ref{w})
\begin{equation}
w=\textstyle{\frac12}(\rho-p)
\end{equation}
where $\rho=-T_t^t$ is the energy density and $p=T_r^r$ the radial pressure. 
The energy flux (\ref{ps}) is
\begin{equation}
\psi=\textstyle{\frac12}(\rho+p)dr.
\end{equation}
Using (\ref{Q0}) one sees that the result will follow if 
\begin{equation}
\rho+p=0.
\end{equation}
This is a known property of nice black holes, e.g.\ \cite{CHNVZ}, one 
derivation going as follows. Any spherically symmetric metric can locally be 
written in the Regge-Wheeler form 
\begin{equation}\label{rw}
ds^2=r^2d\Omega^2+e^{2\varphi}(dr_*^2-dt^2)
\end{equation}
where $(r,\varphi)$ are functions of $(t,r_*)$ and $\varphi$ reduces to the 
Newtonian potential in the Newtonian limit, if $t$ reduces to Newtonian time. 
In static cases, the metric can also be written as 
\begin{equation}
ds^2=r^2d\Omega^2+\left(1-\frac{2m}r\right)^{-1}dr^2-e^{2\varphi}dt^2
\end{equation}
for $\varphi(r)$, $m(r)$, where 
\begin{equation}
1-\frac{2m}r=\left(e^{-\varphi}r'\right)^2
\end{equation} 
and the prime denotes differentiation with respect to $r_*$. A nice black hole 
is defined by 
\begin{equation}
e^{2\varphi}=1-\frac{2m}r
\end{equation}
which gives
\begin{equation}
r'=e^{2\varphi}. 
\end{equation}
One component of the Einstein equations is 
\begin{equation}
4\pi r(\rho+p)=-\left(e^{-2\varphi}r'\right)'
\end{equation}
which gives the result.  

\section{Reissner-Nordstr\"om black holes}
For a Schwarzschild black hole, all the trapped spheres satisfy all the 
refinements. The simplest standard example where this is not so is provided by 
non-extreme Reissner-Nordstr\"om black holes, which are nice black holes 
(\ref{nice}) with 
\begin{equation}
m=M-\frac{Q^2}{2r} 
\end{equation}
where $M$ is the mass and $Q$ the charge, being constants satisfying 
\begin{equation}\label{rn}
M^2>Q^2. 
\end{equation}
As is well known, there are both outer and inner trapping horizons, located at
\begin{equation}
r_\pm=M\pm\sqrt{M^2-Q^2} 
\end{equation}
with trapped spheres for 
\begin{equation}
r_-<r<r_+.
\end{equation}

Since there are inner horizons, one knows in advance that trapped spheres close 
to them will be inner rather than outer trapped. Explicitly, one may calculate 
surface gravity (\ref{kappa}) using the general static formula 
\begin{equation}\label{kappa1}
\kappa=\frac1{2\sqrt{|\gamma|}}\partial_r\left(\sqrt{|\gamma|}\gamma^{rr}\right)
\end{equation}
where $\gamma$ is the normal metric, to find
\begin{equation}
\kappa=\frac{M}{r^2}-\frac{Q^2}{r^3} 
\end{equation}
for which the sign switches at
\begin{equation}
r_o=\frac{Q^2}M. 
\end{equation}
Then $\kappa>0$ for $r>r_o$ to satisfy Definition 2, while $\kappa<0$ for 
$r<r_o$. Elementary analysis using (\ref{rn}) shows that 
\begin{equation}
r_-<r_o<r_+
\end{equation}
so that the switch occurs outside the inner horizon as expected.

Regarding increasingly trapped spheres, one has
\begin{equation}
d\left(\frac1{r^2}\left(\frac{2m}r-1\right)\right)=2\frac{2Q^2-3Mr+r^2}{r^5}dr
\end{equation}
which switches sign at
\begin{equation}
R_\pm=\textstyle{\frac32}\left(M\pm\sqrt{M^2-\textstyle{\frac89}Q^2}\right). 
\end{equation}
Since $k_*\cdot dr=1-2m/r<0$ for trapped surfaces, the desired sign in 
Definition 3 then occurs for $R_-<r<R_+$. Elementary analysis using (\ref{rn}) 
shows that 
\begin{equation}
R_+>r_+ 
\end{equation}
so that all trapped spheres sufficiently close to the outer horizon are 
increasingly trapped, while 
\begin{equation}
r_-<R_-<r_o 
\end{equation}
so the switch occurs inside the inner trapped region. 

In summary, as one starts at the outer horizon $r=r_+$ and moves inwards, the 
spheres are both minimal and outer trapped (by Proposition 6) down to $r=r_o$ 
but not beyond, while still being increasingly trapped down to $r=R_-$ but not 
beyond, which is still outside the inner horizon $r=r_-$. 

\section{Dirty black holes}
Simple examples where all three refinements differ are provided by 
\begin{equation} 
ds^2=r^2d\Omega^2+\left(1-\frac{2M}r\right)^{-1}dr^2
-\left(1-\frac{2M}r\right) \left(1-\frac{L}r\right)^2dt^2 
\end{equation}
where the constants $(M,L)$ are here assumed to satisfy
\begin{equation} \label{ml}
2M>L>0. 
\end{equation}
There is an outer trapping horizon at $r=2M$ and a spatial singularity at 
$r=L$, with trapped spheres for 
\begin{equation}
L<r<2M.
\end{equation}
So the global structure is similar to that of a Schwarzschild black hole, 
except that the singularity has non-zero area. These are evidently dirty black 
holes with mass $M$ and singularity area $4\pi L^2$. For $M\gg L$, they will 
look quite like Schwarzschild black holes from outside. 

The analysis is similar to the above. Regarding increasingly trapped spheres, 
one first finds 
\begin{equation}
m=M 
\end{equation}
then
\begin{equation}
d\left(\frac1{r^2}\left(\frac{2m}r-1\right)\right)=2\frac{r-3M}{r^4}dr
\end{equation}
which gives the desired sign for $R<3M$, by Definition 3. Thus all the trapped 
spheres are increasingly trapped. 

Regarding outer trapped spheres, one finds using (\ref{kappa1}) that
\begin{equation}
\kappa=\frac{(2M+L)r-4ML}{r^2(r-L)}
\end{equation}
for which the sign switches at 
\begin{equation}
r_o=\frac{4ML}{2M+L}. 
\end{equation}
Then $\kappa>0$ for $r>r_o$ to satisfy Definition 2, while $\kappa<0$ for 
$r<r_o$, as before. Elementary analysis using (\ref{ml}) shows that 
\begin{equation}
L<r_o<2M.
\end{equation}
Thus there are inner trapped spheres near the singularity, in distinction to 
the Schwarzschild case. 

Regarding minimal trapped spheres, one can use the Regge-Wheeler form 
(\ref{rw}) with
\begin{equation}
e^{2\varphi}=\left(1-\frac{2M}r\right) \left(1-\frac{L}r\right)^2
\end{equation}
where there is no problem with regarding $\varphi$ as imaginary inside the 
horizon, as it will cancel out in the result. In static cases, one generally 
has 
\begin{equation}
k^ak^b\nabla_a\nabla_br=-k^tk^t\Gamma^*_{tt}\partial_*r
\end{equation}
in terms of the connection coefficient
\begin{equation}
\Gamma^*_{tt}=\partial_*\varphi.
\end{equation}
In this case one finds
\begin{equation}
\partial_*r=\left(1-\frac{2M}r\right)\left(1-\frac{L}r\right)
\end{equation}
and
\begin{equation}
k=\left(1-\frac{L}r\right)^{-1}\partial_t
\end{equation}
so
\begin{equation}
k^ak^b\nabla_a\nabla_br=-\frac12\left(1-\frac{L}r\right)^{-2}
\left(1-\frac{2M}r\right)\partial_r\left(e^{2\varphi}\right)
\end{equation}
where
\begin{equation}
\partial_r\left(e^{2\varphi}\right)=2\frac{(M+L)r^2-(4M+L)Lr+3ML^2}{r^4}
\end{equation}
which gives roots
\begin{equation}
r_m=\frac{3ML}{M+L},\qquad L.
\end{equation}
The desired sign in Definition 1 occurs for $r>r_m$. Elementary analysis using 
(\ref{ml}) shows that 
\begin{equation}
r_o<r_m<2M
\end{equation}
so the switch occurs in the outer trapped region.

In summary, as one starts at the horizon $r=2M$ and moves inwards, the spheres 
are minimal trapped down to $r=r_m$ but not beyond, outer trapped down to 
$r=r_o$ but not beyond, while still being increasingly trapped all the way to 
the singularity $r=L$. 

For completeness, one may note that the energy density vanishes,
\begin{equation}
\rho=0
\end{equation}
which is a general property if $m$ is constant, and that the radial pressure is 
\begin{equation}
p=\frac{(M+L)(r-r_m)}{2\pi r^3(r-L)}.
\end{equation}
The dominant energy condition requires $\rho\ge p$, which is violated for 
$r>r_m$. On the other hand, it is satisfied for $r\le r_m$ and so the hierarchy 
of trapped surfaces is the standard one according to Propositions 1--2. 

Since $p$ is rapidly vanishing as $r\to\infty$, $O(r^{-3})$, but infinite as 
$r\to L$, one can interpret it as coming from some modification of Einstein 
gravity at small length scales. The scale is given by $L$, which might be the 
Planck length or some other theoretical scale. 

\section{Summary}
The hierarchy of trapped spheres is illustrated as follows:
\begin{eqnarray}
&&\hbox{involute}\\
&&\quad\Downarrow\nonumber\\
&&\hbox{minimal}\nonumber\\
&&\quad\Downarrow\nonumber\\
&&\hbox{somehow involute}\Rightarrow\hbox{outer}\Rightarrow\hbox{anyhow increasingly}\nonumber\\
&&\qquad\qquad\qquad\quad\hbox{NEC}\qquad\hbox{NEC}\qquad\Downarrow\nonumber\\
&&\qquad\qquad\qquad\qquad\qquad\qquad\qquad\hbox{increasingly}\nonumber\\
&&\qquad\qquad\qquad\qquad\qquad\qquad\qquad\qquad\Downarrow\nonumber\\
&&\qquad\qquad\qquad\qquad\qquad\qquad\hbox{somehow increasingly}\nonumber
\end{eqnarray}
where the vertical implications are straightforward, while the horizontal 
implications require NEC. In general cases, the vertical hierarchy will remain, 
but the horizontal hierarchy will change and be interwoven \cite{ts2}. 

In static cases, the vertical hierarchy collapses to leave just
\begin{eqnarray}
&&\hbox{minimal}\Rightarrow\hbox{outer}\Rightarrow\hbox{increasingly}\\
&&\qquad\quad\hbox{NEC}\qquad\hbox{NEC}\nonumber
\end{eqnarray}
and the first two become equivalent for the widely considered subclass of nice 
(non-dirty) black holes. Examples have also been given of dirty black holes 
where all three refinements differ. The examples show that the refinements can 
still hold sufficiently close to the outer horizon of a black hole, even if it 
is dirty and violates energy conditions. On the other hand, the refinements 
need not hold near an inner horizon or a singularity. This provides some 
support for the idea to restrict trapped surfaces to those which are regular 
enough near an outer trapping horizon that the horizon is indeed the boundary 
of the region of such refined trapped surfaces. 

Finally, while geometrical ideas play a large part in the above definitions and 
results, recall that the original motivation was a physical one, namely to have 
a practical definition of black holes and a comprehensive theory. One might 
therefore argue that perhaps the most physically well motivated refinement is 
outer trapped, since it expresses positivity of surface gravity $\kappa$, which 
appears in a first law for trapping horizons \cite{1st}, determines a local 
Hawking temperature $\kappa/2\pi$ \cite{HCNVZ} and has other physically 
relevant properties \cite{1st,ine}. In any case, the interplay of geometry and 
physics remains a fascinating aspect of black holes, 95 years after the 
Schwarzschild solution was found. 

\medskip
Research supported by the National Natural Science Foundation of China under 
grants 10375081, 10473007 and 10771140, by Shanghai Municipal Education 
Commission under grant 06DZ111, and by Shanghai Normal University under grant 
PL609.


\begin{thebibliography}{99}
\bibitem{Pen}R Penrose, {Phys. Rev. Lett.} {\bf 14}, 57 (1965)
\bibitem{Haw}S W Hawking, {Proc. R. Soc. London} {\bf A300}, 187 (1967) 
\bibitem{HP}S W Hawking \& R Penrose, {Proc. R. Soc. London} {\bf A314}, 529 
    (1970) 
\bibitem{AK}A Ashtekar \& B Krishnan, {Living Rev. Relativity} {\bf 7}, 10 
    (2004)
\bibitem{Boo}I Booth, {Can. J. Phys.} {\bf 83}, 1073 (2005)
\bibitem{Kri}B Krishnan, {Class. Quantum Grav.} {\bf 25}, 114005 (2008)
\bibitem{GJ}E Gourgoulhon \& J L Jaramillo, {New Astron. Rev.} {\bf 51}, 791 
    (2008)
\bibitem{bh6}S A Hayward, {Adv. Sci. Lett.} {\bf 2}, 205 (2009) 
\bibitem{SK}E Schnetter \& B Krishnan, {Phys. Rev.} {\bf D73}, 021502 (2006) 
\bibitem{Ben}I Ben-Dov, {Phys. Rev.} {\bf D75}, 064007 (2007)
\bibitem{BS}I Bengtsson \& J M M Senovilla, {Phys. Rev.} {\bf D79}, 024027 
    (2009)  
\bibitem{1st}S A Hayward, {Class. Quantum Grav.} {\bf 15}, 3147 (1998)
\bibitem{HCNVZ}S A Hayward, R Di Criscienzo, L Vanzo, M Nadalini \& S Zerbini, 
    {Class. Quantum Grav.} {\bf 26}, 062001 (2009) 
\bibitem{New}R P A C Newman, {Class. Quantum Grav.} {\bf 4}, 277 (1987)
\bibitem{KH}M Kriele \& S A Hayward, {J. Math. Phys.} {\bf 38}, 1593 
    (1997) 
\bibitem{AMS}L Andersson, M Mars, \& W Simon, {Phys. Rev. Lett.} {\bf 95} 
    111102 (2005) 
\bibitem{AMMS}L Andersson, M Mars, J Metzger \& W Simon, {Class. Quantum Grav.} 
    {\bf 26}, 085018 (2009)  
\bibitem{ts2}S A Hayward, Involute, minimal, outer and increasingly trapped 
    surfaces (arXiv:0906.2528) 
\bibitem{Vis}M Visser, {Phys. Rev.} {\bf D46}, 2445 (1992); {Phys. Rev.} {\bf 
    D48}, 583 (1993) 
\bibitem{sph}S A Hayward, {Phys. Rev.} {\bf D53}, 1938 (1996) 
\bibitem{ine}S A Hayward, {Phys. Rev. Lett.} {\bf 81}, 4557 (1998) 
\bibitem{Kod}H Kodama, {Prog. Theor. Phys.} {\bf 63}, 1217 (1980)
\bibitem{MHC}H Maeda, T Harada \& B J Carr, {Phys. Rev.} {\bf D79}, 044034 
    (2009) 
\bibitem{MS}C W Misner \& D H Sharp, {Phys. Rev.} {\bf 136}, B571 (1964)
\bibitem{CHNVZ}R Di Criscienzo, S A Hayward, M Nadalini, L Vanzo \& S Zerbini, 
    arXiv:0906.1725
\end{thebibliography}
\end{document}